\documentclass[%
 reprint,
 amsmath,amssymb,
 aps,
]{revtex4-1}

\usepackage{graphicx}
\usepackage{svg}
\usepackage[caption=false]{subfig}
\usepackage{svg}
\usepackage{dcolumn}
\usepackage{bm}
\usepackage{physics}
\usepackage{mathtools}
\usepackage{enumerate}

\usepackage{comment}
\excludecomment{outlinesec}
\usepackage[normalem]{ulem}

\begin{document}

\title{Entropic Efficiency of Bayesian Inference Protocols}

\author{Nathan Shettell}
\author{Alexia Auffèves}%

\affiliation{%
 MajuLab, CNRS-UCA-SU-NUS-NTU International Joint Research Laboratory
}%
\affiliation{%
 Centre for Quantum Technologies, National University of Singapore, 117543 Singapore, Singapore
}%

\newcommand{\nate}[1]{\textcolor{blue}{#1}}
\newcommand{\red}[1]{\textcolor{red}{#1}}
\newcommand{\natestrike}[1]{\textcolor{red}{\sout{#1}}}

\date{\today}

\begin{abstract}
    Inference is a versatile tool that underlies scientific discovery, machine learning, and everyday decision-making: it describes how an agent updates a probability distribution as partial information is acquired from multiple measurements, reducing ignorance about a system’s latent state. We define an inferential efficiency as the ratio of information gain to cumulative memory erasure cost, with inefficiency arising from unexploited correlations between the measured system and memories, and/or between memories and environment (noise). Using this efficiency, we benchmark two limiting measurement paradigms: sequential, in which the same memory is exploited iteratively, and parallel, in which many memories are exploited simultaneously. In both cases, the minimal erasure cost reflects correlations across memories: temporal in sequential, spatial in parallel. Remarkably, when all system-memory correlations are exploited for inference, both paradigms attain the same minimal erasure cost, even in the presence of noise. Conversely, the parallel paradigm performs better in the presence of unexploited correlations, stemming from hidden memories' degrees of freedom. This approach provides a quantitative, physically grounded criterion to compare inference strategies, determine their efficiency, and link target information gains to their minimal entropic cost.
\end{abstract}

\maketitle


Inference, in its mathematical formulation, is a process of updating a probability distribution in light of new data: it is a classical computation that reduces an agent's ignorance about a system’s latent state \cite{kay1993fundamentals, cover1999elements, hastie2009elements, sivia2006data, cox2006principles}. This process underlies applications ranging from cosmology \cite{ramanah2019cosmological} to genomics \cite{zhu2017bayesian}, and  forms the foundations of modern machine learning \cite{murphy2012machine}. As model and dataset scales continue to increase, the energetic costs associated with these inference steps have become a central concern, prompting interest in the physical limits of information processing \cite{strubell2019energy, aquino2025towards, samsi2023words, mavromatis2024computing}. These limits reflect that inference is fundamentally constrained by the physics of information, as it relies on the generation of system–memory correlations (the \emph{information}) during measurement \cite{landauer1961irreversibility, maruyama2009colloquium, sagawa2009minimal, parrondo2015thermodynamics, minagawa2025universal}. These correlations can be exploited to progressively reduce the system’s entropy quantifying the agent's ignorance through an inference process, but this reduction is not free: after the measurement, the state of the memory itself has become uncertain, which is captured by an increase of its entropy. This entropy sets a baseline for the fundamental cost of erasure, and hence the entire process. These entropy flows are governed by the same thermodynamic principles which underlie measurement, feedback and erasure in a Maxwell's demon experiment \cite{parrondo2015thermodynamics, minagawa2025universal, lubkin1987keeping, maroney2009information,sagawa2008second, sagawa2009minimal}, in which knowledge gain replaces work extraction \cite{saha2021maximizing, song2021optimal}. \\

Considering this perspective, we analyze inference from a purely entropic viewpoint, providing an operationally agnostic framework to compare different measurement strategies. We focus on Bayesian protocols, in which a prior over the system state is updated via a well-defined likelihood function, complementing established links between Bayesian statistics and thermodynamics \cite{buscemi2021fluctuation, aw2021fluctuation}. Our approach is, however, general and can be extended to non-Bayesian or learning-based inference schemes \cite{vapnik1999overview, goodfellow2016deep}. In Bayesian protocols, individual measurements typically yield only partial information about the system \cite{still2020thermodynamic}, due either to imperfect measurement or to structured memories in which only a subset of degrees of freedom are accessible for inference; necessitating iterative updates. These iterated cycles generate correlations across memory registers, whose nature (temporal or spatial) defines distinct implementations. The impact of such correlations has been highlighted in the performances of predictive inference \cite{still2012thermodynamics, still2020thermodynamic, boyd2025thermodynamic} and modular computation \cite{boyd2018thermodynamics, riechers2018transforming}. Here, we explore how leveraging these correlations impacts the cumulative erasure cost relative to treating each memory independently, hence the efficiency of the whole protocol. This Letter is organized as follows. We begin with an entropic analysis of an autonomous single-cycle measure–infer–erase protocol. We then extend this framework to multiple measurements, contrasting sequential reuse of a single memory with a parallel architecture employing multiple memories simultaneously. An explicit example of inferring a classical bit using a structured memory illustrates the resulting efficiencies.


\begin{figure*}
    \centering
    \includegraphics[width=0.85\linewidth]{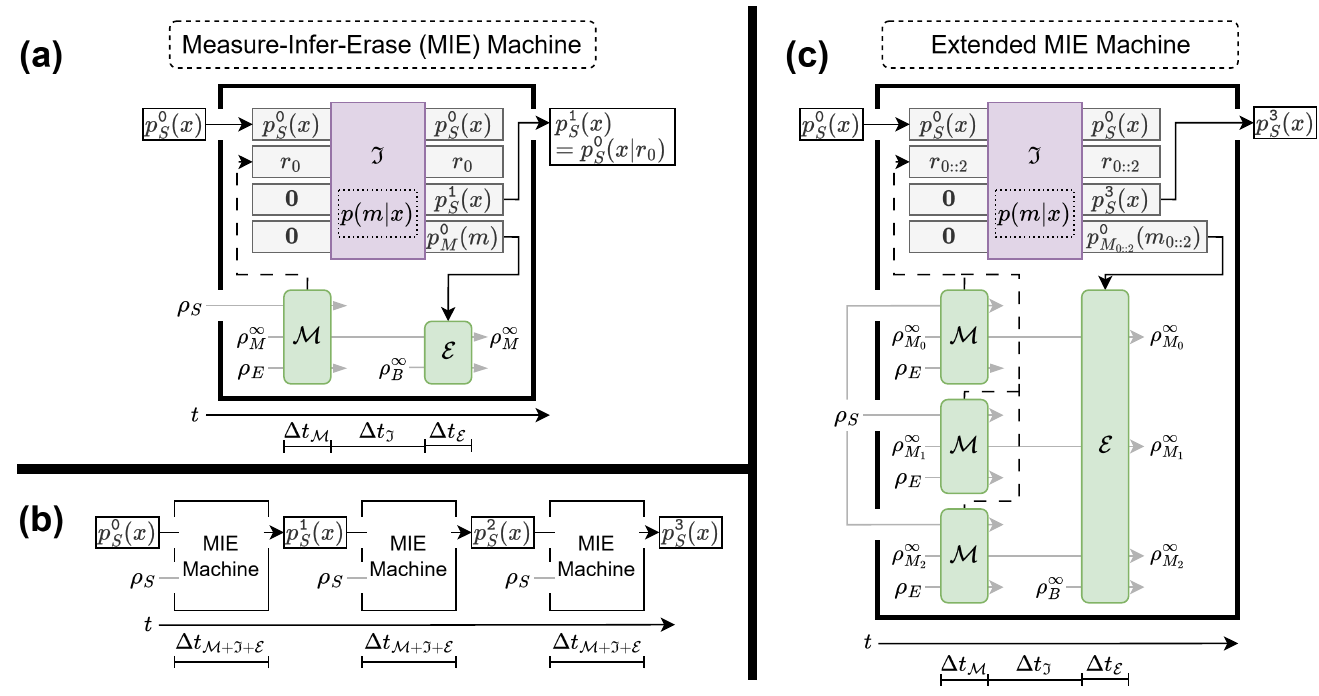}
    \caption{(a) Autonomous measure–infer–erase cycle. An agent provides the system $S$ together with a prior $p_S^\mathtt{0}(x)$, which is processed by an autonomous machine equipped with a structured memory $M=(Q,R)$, and environment $E$, and a thermal bath $B$.  Measurement via an entropy-preserving map $\mathcal{M}$ generates correlations in $Q$ and a coarse-grained outcome $r_0$ in $R$, which is then processed by a reversible Bayesian update to yield $p_S^\mathtt{1}(x)$. An optimal erasure protocol subsequently resets the memory through coupling to the thermal bath. (b) Sequential architecture ($n=3$), in which the measure-infer-erase cycle is iterated: the posterior from each cycle becomes the prior for the next, enabling temporal correlations during erasure. (c) Parallel architecture ($n=3$), realized by extending the autonomous machine to include multiple memories that record outcomes concurrently and are erased simultaneously, enabling spatial correlations during erasure. Together, the sequential and parallel paradigms exhibit complementary trade-offs between temporal resources and hardware complexity.}
    \label{fig:InferenceCycle}
\end{figure*}

\emph{Single Cycle \textemdash} We first focus on a single iteration of an autonomous measure-infer-erase cycle, depicted in Fig~(\ref{fig:InferenceCycle}a). An agent designs a machine that takes as input the unknown state of the system, $\rho_S$, together with a prior distribution $p_S^\mathtt{0}(x)$, and outputs an updated distribution $p_S^\mathtt{1}(x)$, where $x$ indexes a specific possible system state. The machine is equipped with a structured memory $M = (Q , R)$, where $Q$ ($R$) represents inaccessible (accessible) degrees of freedom, $Q$ correlating with the system, and $R$ correlating with $Q$. In this view, $R$ is the register that encodes the \emph{useful} measurement outcomes exploited for inference. The possible incomplete transmission of information from $Q$ to $R$ is a natural source of partiality \cite{still2020thermodynamic}, a specific example of this construction is to take $R$ as a coarse-gaining of $Q$. This decomposition provides a flexible model for a variety of measurement schemes. Additional components include a bath $B$ in thermal equilibrium $\rho_B^\infty$ for memory erasure \cite{ThermalNote}, an environment $E$ responsible for the noise affecting the measurement ($E$ is assumed to be large and weakly coupled, such that its state remains approximately unperturbed), and a reversible computer \cite{bennett1973logical} which outputs the updated distribution. Programming this computer requires the agent to have a physical model of system-memory interaction (see below). Together, these constituents allow the entire cycle to be modeled as an entropy-preserving map. In what follows, we detail the three steps of the cycle, tracking the flows of entropy. All entropies are informational, i.e. they capture the agent's ignorance about the system and the memory state.\\

\noindent (1) \emph{Measurement} ($t \in \Delta t_\mathcal{M}$) \textemdash The system $S$ interacts with the memory $M$, which is initially prepared in the thermal equilibrium state $\rho_M^\infty$, and the environment $E$, all initially uncorrelated. The measurement is assumed to be non-demolition, such that the marginal of $S$ remains unchanged, while correlations are established between $S$, $M$ and $E$. The joint evolution is modeled by an entropy-preserving map $\mathcal{M}$. Tracing over environmental degrees of freedom yields the system-memory distribution
\begin{equation}
    p_{SM}^\mathtt{0}(x,m)=p_{SQR}^\mathtt{0}(x,q,r),
\end{equation}
where $m, q, r$ index specific states of $M, Q, R$ respectively. The superscript $\mathtt{0}$ labels quantities relative to the current prior, and is incremented in subsequent cycles to track successive updates of the agent’s knowledge. The measurement produces a specific outcome $r_0$, which serves as the information input to the subsequent inference step.\\

\noindent (2) \emph{Inference} ($t \in \Delta t_\mathfrak{I}$) \textemdash The agent updates their description of the system using Bayes' rule, conditioned on the outcome $r_0$:
\begin{equation}
    \label{eq:BayesUpdate}
    p_S^\mathtt{1}(x)=p_S^\mathtt{0}(x|r_0)= \frac{p^\mathtt{0}_S(x)p(r_0|x)}{p_R^\mathtt{0}(r_0)}.
\end{equation}
The likelihood $p(r_0|x)$ is derived from a physical model of measurement, while $p_R^\mathtt{0}(r_0)=\sum_x p_S^\mathtt{0}(x) p(r_0|x)$ represents the agent’s guess for observing the outcome $r_0$. The gain of knowledge provided by the update is captured by the decrease of the system's entropy,
\begin{equation}
    \mathcal{I}=H(p_S^\mathtt{0} )- H(p_S^\mathtt{1} )= I(p_S^\mathtt{0}:p_R^\mathtt{0}).
\end{equation}
$I(p_S^\mathtt{0}:p_R^\mathtt{0})$ is the mutual information in the post-measurement joint distribution $p_{SR}^\mathtt{0}(x,r)$, where the mutual information $I(p_A:p_B)$ is generally defined as
\begin{equation}
    I(p_A:p_B)=\sum_{a,b}p_{AB}(a,b)\log_2 \frac{p_{AB}(a,b)}{p_A(a)p_B(b)}.
\end{equation}
By construction, the Bayesian update captured by Eq.~\eqref{eq:BayesUpdate} conserves the joint entropy of $(S,R)$. This is consistent with the fact that it can be implemented via reversible computation, incurring no entropic cost. Reversible computing  \cite{bennett1973logical} requires to conserve the inputs and to store the outputs on ancillary registers, see Fig.~(\ref{fig:InferenceCycle}a). In addition to $p^\mathtt{1}_S(x)$, the computer outputs the memory distribution $p_M^\mathtt{0}(m)=\sum_xp^\mathtt{0}_{SM}(x,m)$ required for the subsequent erasure step. The marginal used for the update, $p_R^\mathtt{0}(r)$, is simply the result of tracing over $Q$, underscoring that only part of the correlations can be exploited for inference.\\

\noindent (3) \emph{Smart Erasure} ($t \in \Delta t_\mathcal{E}$) \textemdash
To complete the cycle, the memory is autonomously reset to its initial state, $\rho_M^\infty$ by a pre-programmed erasure operation $\mathcal{E}$ that couples $M$ to a thermal bath $B$. As shown in Ref.~\cite{rio2011thermodynamic}, the minimal erasure cost is determined by the agent’s prior knowledge on the memory state, here encoded in the distribution $p_M^\mathtt{0}(m)$:
\begin{equation}
    \mathcal{C}_\mathtt{0} =\Delta H_M=H (p_M^\mathtt{0} )-H (\rho_M^\infty ).
\end{equation}
Achieving this limit requires the design of \emph{smart} erasure protocols that exploit the knowledge $p_M^\mathtt{0}$ supplied by the inference step to optimally remove the entropy stored in the memory, see Suppl. Mat.~\ref{supp_mat:erasure} for an explicit single-bit example.\\

Eventually, the three steps (measurement, inference, and erasure) form a process in which system–memory correlations are created and exploited, before the memory is erased. This perspective naturally motivates an entropic notion of inferential efficiency, defined as the ratio of information gained to erasure cost
\begin{equation}
\label{eq:eff1}
    \eta = \frac{\mathcal{I}}{\mathcal{C}_\mathtt{0}} 
    = 1-\frac{\Delta H_{SM}+I(p_{S}^\mathtt{0}:p_{M}^\mathtt{0})-I(p_{S}^\mathtt{0}:p^\mathtt{0}_R)}{\Delta H_M},
\end{equation}
where the final expression follows from the non-demolition nature of the measurement. By construction, $\eta \leq 1$, with equality only when all entropy transferred into the memory corresponds to system-memory correlations usable for inference. Two mechanisms reduce efficiency in practice. First, $\Delta H_{SM}$ quantifies entropy injected into the system-memory pair due to correlations created with the environment (noise). Second, the difference $I(p_{S}^\mathtt{0}:p_{M}^\mathtt{0})-I(p_{S}^\mathtt{0}:p^\mathtt{0}_R)$ captures correlations between $S$ and $Q$ that are not reflected within $R$, therefore cannot be exploited for inference. Together, these contributions represent the true cost of irreversibility~\cite{parrondo2015thermodynamics}, i.e. the erasure of remnant correlations which were not leveraged. The resulting efficiency $\eta$ provides a physically grounded criterion for comparing inference strategies on equal entropic footing. \\


\emph{Multiple Cycles \textemdash} Practical inference requires multiple measurements to progressively improve the agent’s knowledge. We therefore introduce a collection of $n$ memories, denoted $M_0,\ldots,M_{n-1}$ as well as the joint prior probability distribution over the system and memory registers
\begin{equation}
\label{eq:multi_distribution}
    p^\mathtt{0}_{S,M_0,\ldots,M_{n-1}}(x,m_0,\ldots,m_{n-1})=p^\mathtt{0}_S(x) \prod_{k=0}^{n-1} p(m_k|x).
\end{equation}
Let us now consider the two implementations of the cumulative inference process depicted in Fig.~(\ref{fig:InferenceCycle}b,c).
In the \emph{sequential} implementation, a single physical memory is reused across iterated cycles, Fig.~(\ref{fig:InferenceCycle}b). There, $M_k$ refers to the memory when the $(k+1)^{th}$ iteration starts, and thus the joint-distribution instead captures temporal correlations across successive memory states. This architecture minimizes hardware requirements but requires information processing to take place over longer operational timescales. By contrast, in the \emph{parallel} implementation, the registers $M_0,\ldots,M_{n-1}$ denote distinct physical memories that are simultaneously correlated with the system during a single measurement stage, Fig.~(\ref{fig:InferenceCycle}c). In this case, the joint distributions describes spatial correlations across the memories which can be exploited as the cost of a hardware overhead. \\

Whichever the protocol, the inference step and the agent's resulting posterior distribution, depend only on the joint statistics of the system and memories $p^\mathtt{n}_S(x)=p^\mathtt{0}_S(x|r_0,\ldots,r_{n-1})$, where $r_k$ denotes for the measurement outcome of the $(k+1)^{th}$ cycle. Hence, the cumulative information gain does not depend on the implementation,
\begin{equation}
    \mathcal{I}(n)=H(p_S^\mathtt{0})-H(p_S^\mathtt{n})=I(p_{S}^\mathtt{0}:p_{R_{0::n-1}}^\mathtt{0}),
\end{equation}
where
\begin{equation}
    p_{R_{0::n-1}}^\mathtt{0}(r_0,\ldots,r_n)=\sum_xp^\mathtt{0}_S(x) \prod_{k=0}^{n-1} p(r_k|x)
\end{equation}
denotes the joint distribution of states characterizing the collection of accessible registers $R_0,\ldots,R_{n-1}$. By contrast, the minimal cost of erasing all 
$n$ memories \emph{does} depend on how cross-correlations can be processed during erasure. We distinguish three erasure costs associated with the joint distribution in Eq.~\eqref{eq:multi_distribution}: a baseline cost that ignores all cross-memory correlations, and the minimal costs associated with the parallel and sequential implementations. The baseline corresponds to an \emph{uncorrelated} erasure strategy, in which each memory is optimally reset independently. Its minimal cost is
\begin{equation}
\label{eq:basecost}
    \mathcal{C}_\otimes (n) = \sum_{k=0}^{n-1} \big( H (p_{M_k}^\mathtt{0})-H(\rho_{M}^\infty) \big)=n \mathcal{C}_\mathtt{0}.
\end{equation}
In the \emph{parallel} implementation, all $n$ memories are erased simultaneously. The minimal cumulative cost is
\begin{equation}
\begin{split}
    \mathcal{C}_\text{par} (n) &=H(p_{M_{0::n-1}}^\mathtt{0})-nH(\rho_{M}^\infty) \\
    &=\mathcal{C}_\otimes (n)-\sum_{k=1}^{n-1}I(p_{M_k}^\mathtt{0}:p_{M_{0::k-1}}^\mathtt{0}),
\end{split}
\end{equation}
where $p^\mathtt{0}_{M_{0::k-1}}$ is the marginal of the joint distribution in Eq.~\eqref{eq:multi_distribution} over the memories $M_0,\ldots,M_{k-1}$. Conversely in the \emph{sequential} implementation, correlations are accessed only through past measurement outcomes, yielding
\begin{equation}
\begin{split}
    \mathcal{C}_\text{seq} (n) &= \sum_{k=0}^{n-1} \big( H(p^\mathtt{0}_{M_k|R_{0::k-1}})-H(\rho_{M}^\infty) \big) \\
    &=\mathcal{C}_\otimes (n)-\sum_{k=1}^{n-1}I(p^\mathtt{0}_{M_k}:p^\mathtt{0}_{R_{0::k-1}}),
\end{split}
\end{equation}
where $H(p_{M_k|R_{0::k-1}})$ denotes the conditional entropy of $M_k$ given the states of $R_0,\ldots,R_{k-1}$. A visual derivation of the entropy landscape underlying these costs, illustrated for the base case 
$n=2$, is provided in the Suppl. Mat.~\ref{supp_mat:venn}. \\

The mutual-information terms in $\mathcal{C}_\text{par}$ and $\mathcal{C}_\text{seq}$ represent \emph{discounts} relative to the baseline cost $\mathcal{C}_\otimes$. Specifically, they quantify how correlations can be leveraged during erasure to reduce the cumulative cost; spatial correlations in the parallel implementation, and temporal correlations in the sequential implementation. Since each memory decomposes as $M_k = (Q_k, R_k)$, one always has
\begin{equation}
    I(p^\mathtt{0}_{M_k}:p^\mathtt{0}_{M_{0::k-1}}) \geq I(p^\mathtt{0}_{M_k}:p^\mathtt{0}_{R_{0::k-1}}) \;\Rightarrow\mathcal{C}_\text{par} \leq \mathcal{C}_\text{seq}.
\end{equation}
Remarkably, the two costs coincide when $H(p_{M_k}^\mathtt{0})=H(p_{R_k}^\mathtt{0})$, i.e. when all system–memory correlations are exploitable for inference, even in the presence of noise. In this case, spatial and temporal correlations are aligned, and the remaining trade-off lies solely between temporal overhead in sequential implementations and hardware cost in parallel architectures. Otherwise, the excess cost of the sequential strategy reflects a cumulative penalty incurred by erasing memories using only a partial set of correlations: those encoded in $R$ but not in $Q$. Importantly, this distinction is independent of environmental noise and arises solely from the structure of information transmission within each memory.\\

Because $\mathcal{I}(n)$ is independent of measurement implementation, differences in entropic efficiency arise from how correlations are stored, propagated and ultimately erased. This yields the natural hierarchy
\begin{equation}
    \label{eq:hierarchy}
    \mathcal{I}(n)\leq \mathcal{C}_\text{par} (n) \leq \mathcal{C}_\text{seq} (n) \leq \mathcal{C}_\otimes (n),
\end{equation}
which generates a decreasing ladder of inferential efficiency, $\eta(n) = \mathcal{I}(n)/\mathcal{C}(n)$. The upper bound is attained only when all memory entropy corresponds to correlations that are fully exploitable for inference, representing perfect alignment between information gained and entropy expended.\\


\begin{figure*}
    \centering
    \includegraphics[width=0.95\linewidth]{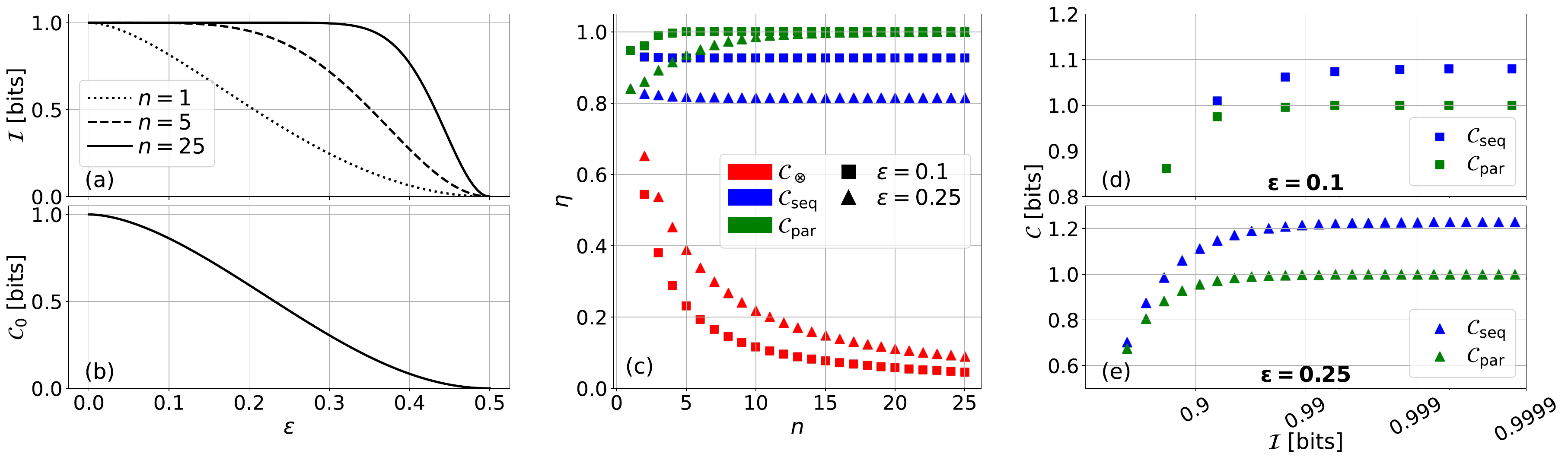}
    \caption{Illustrative example of inferential efficiency for multiple measurements of a binary system. (a) The information gain $\mathcal{I}(n)$ increases rapidly for small $\varepsilon$, reaching near-maximal certainty for modest $n$. (b) The cost of a single-cycle $\mathcal{C}_0$, similarly increases as $\varepsilon$ decreases; crucially $\mathcal{C}_0 > \mathcal{I}(1)$ for all $\varepsilon \in (0,1/2)$, reflecting the inefficiency arising from unexploited system-memory correlations. (c) Entropic efficiency of the three paradigms as a function of $n$: parallel approaches unity, sequential plateaus to a finite value, and uncorrelated decays to zero. (d,e) Information gain versus erasure cost (metric-resource representation) for sequential and parallel strategies at $\varepsilon=0.1$ and $\varepsilon=0.25$ , respectively, illustrating the widening efficiency gap as noise increases.}
    \label{fig:ex}
\end{figure*}

\emph{Example \textemdash} To illustrate the distinction between sequential and parallel measurement paradigms, we consider inferring the state of a classical bit $S$ ($x\in \{0,1 \}$) using a four-bit memory $M=(Q,R)$, where $Q$ consists of three binary degrees of freedom and $R$ is an accessible register that records a majority vote. Each of the three bits in $Q$ interacts independently and identically with the system during measurement, while $R$ deterministically encodes the majority value of their post-measurement states. Imperfect preparation of the memory yields the initial state
\begin{equation}
\rho_M^\infty =
\underbrace{ \big( (1-\varepsilon)\dyad{g_c}
+\varepsilon \dyad{g_i} \big)^{\otimes 3} }_{\rho_Q^\infty}
\;\otimes\;
\underbrace{\dyad{\boldsymbol{0}}}_{\rho_R^\infty}.
\end{equation}
with entropy $H(\rho_M^\infty)=3h(\varepsilon)$, where $h$ is the binary entropy function. During measurement, $\ket{g_c}$ and $\ket{g_i}$ respectively record the system state correctly or incorrectly, with $\varepsilon$ controlling the initial noise in the inaccessible memory. This example contains no environmental noise, so that inefficiency arises solely from the incomplete transmission of information from $Q$ to $R$, see Eq.~\eqref{eq:eff1}. All entropies and information measures are expressed in bits; detailed calculations are provided in Suppl. Mat.~\ref{supp_mat:example}.\\

Fig.~(\ref{fig:ex}) summarizes the information gain and erasure costs for the three measurement paradigms. The information gain increases rapidly with the number of measurements, with the agent’s uncertainty decaying exponentially as successive outcomes refine the posterior, Fig.~(\ref{fig:ex}a). Consequently, even for moderate preparation noise, only a small number of measurements is sufficient to approach near-maximal certainty. The associated erasure costs, however, differ noticeably across strategies. For the uncorrelated strategy, independent erasure makes the per-cycle cost largest for a perfectly prepared memory and monotonically smaller with increasing noise $\varepsilon$, Fig.~(\ref{fig:ex}b). This leads to a counterintuitive behavior: the overall efficiency $\eta_\otimes$ increases with $\varepsilon$ despite reduced information gain, but decreases with the number of measurements as erasure costs accumulate linearly while the information gain saturates, Fig.~(\ref{fig:ex}c). By contrast, strategies that exploit correlations do not exhibit this anomalous behavior. In the parallel implementation, the cumulative erasure cost converges faster than the information gain, so the efficiency increases monotonically with the number of measurements and asymptotically approaches unity. The sequential strategy lies between these extremes: erasing each memory before the next measurement leaves an overhead, dominated by the first few cycles, so the efficiency saturates below the parallel limit. As a result, its efficiency exceeds that of the uncorrelated strategy but saturates at a finite plateau below the parallel limit, Fig.~(\ref{fig:ex}c). \\

While the number of measurements $n$ provides a convenient operational handle, it is often more physically informative to examine the information gain as a function of the corresponding erasure cost, as shown in Fig.~(\ref{fig:ex}d,\ref{fig:ex}e). In such a metric–resource representation, one can determine either the maximal information attainable for a fixed entropy budget or the erasure cost required to reach a target information level. In the present example, both sequential and parallel strategies achieve any desired information gain with finite cost, but the gap between them widens with increasing $\varepsilon$, highlighting the growing advantage of exploiting spatial correlations in noisier regimes. This viewpoint clarifies the operational trade-off between temporal overhead in sequential implementations and hardware overhead in parallel ones, directly linking entropic constraints to broader resource considerations. \\


\emph{Discussion \textendash} In this work, we assign explicit physical operations to inference, framing it as a cycle of measurement, update and erasure. At the single-cycle level, this is a realization of Maxwell's demon that converts ignorance into structured knowledge rather than work. We introduce a natural notion of inferential efficiency, defined as the ratio of epistemic gain to the cost of erasure, which reaches unity only when all memory entropy stems from system ignorance; inefficiency arises from inexploitable system-memory or system-environment correlations. While we formulate the framework using Bayesian updating with known likelihoods $p(m|x)$ and system priors $p_S^\mathtt{0}(x)$, it naturally extends to other update rules and information-theoretic metrics \cite{vapnik1999overview, goodfellow2016deep}.\\

In the multiple-measurement regime, two limiting paradigms arise: sequential and parallel; distinguished by whether temporal or spatial correlations across memories are leveraged. While both paradigms ultimately yield the same inferred state, they differ in how efficiently they can recycle correlations during erasure. In the absence of inexploitable system-memory correlations for inference, their inferential efficiency converges, and the choice between them is governed by practical constraints such as hardware overhead and timescale of physical operations.  These results offer a quantitative foundation for optimizing inference architectures, principles which underpin methodologies across science, engineering and modern computation alike. Immediate applications include inference-intensive tasks such as metrology \cite{degen2017quantum} and tomography \cite{o2016efficient}, as well as contemporary machine learning \cite{murphy2012machine}, where the energetic cost of inference can represent a significant bottleneck \cite{strubell2019energy, aquino2025towards, samsi2023words, mavromatis2024computing}.  \\




\textbf{Acknowledgments \textendash} This project is supported by the National Research Foundation, Singapore through the National Quantum Office, hosted in A*STAR, under its Centre for Quantum Technologies Funding Initiative (S24Q2d0009). The authors acknowledge the Plan France 2030 through the project BACQ as part of the MetriQs-France program (Grant ANR-22-QMET-0002) and through the project OECQ funded by BPIFrance. AA thanks M.Richard for insightful discussions.

\bibliographystyle{unsrt}
\bibliography{main}



\newpage 

\setcounter{figure}{0}
\renewcommand{\thefigure}{S\arabic{figure}}

\setcounter{equation}{0}
\renewcommand{\theequation}{S\arabic{equation}}

\setcounter{section}{0}
\renewcommand{\thesection}{S\arabic{section}}
\renewcommand{\thesubsection}{S\arabic{section}.\arabic{subsection}}

\widetext
\newpage

\begin{center}
\vskip0.5cm
{\Large Supplemental Material}
\end{center}

\setcounter{equation}{0}
\setcounter{figure}{0}
\setcounter{table}{0}
\setcounter{section}{0}

\renewcommand{\theequation}{S\arabic{equation}}
\renewcommand{\thefigure}{S\arabic{figure}}
\renewcommand{\thesection}{S\Roman{section}}

\section{Smart of Erasure of a Bit} \label{supp_mat:erasure}

In the following, we examine how an agent can optimally reset a binary memory $M$ to its initial state $\rho_M^\infty$. Consider two scenarios: in the first, the agent’s description of the memory is given by the uniform distribution $p_M(m)=\frac{1}{2}$ for $m \in \{0,1\}$; in the second, the memory is described by a biased distribution $p_M^\prime(m)=\frac{1\pm(-1)^m \delta}{2}$, where $\delta$ quantifies the exploitable bias. Knowing the respective distribution, the agent can design an erasure operation, $\mathcal{E}[p_M]$ or $\mathcal{E}[p_M^\prime]$, to reset the memory efficiently at minimal entropic cost. Crucially, $\mathcal{E}$ is outcome-independent: it does not depend on the realized memory state, but is tailored to the known statistics of the memory. In the biased case, this prior knowledge allows the agent to implement an erasure protocol that removes only the entropy actually present, thereby enabling a lower-cost reset.\\

\begin{figure}[h]
    \centering
    \includegraphics[width=0.75\textwidth]{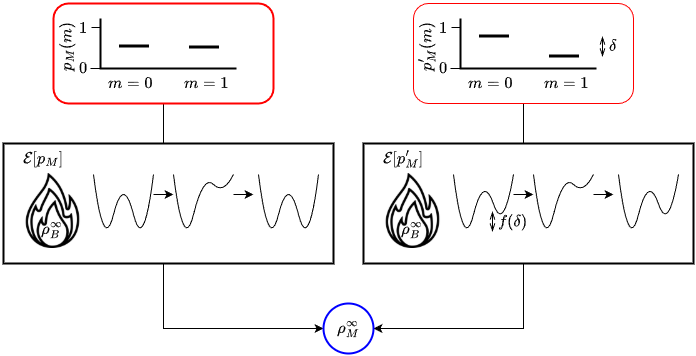}
    \caption{Erasure operations optimized for distinct memory statistics, implemented via coupling to a thermal bath and leading to the same reset state with different minimal costs.}
    \label{fig:smart_erasure}
\end{figure}

Fig.~(\ref{fig:smart_erasure}) illustrates this principle. The first row shows the initial distributions, $p_M(m)$ and $p_M^\prime(m)$. The second row schematically depicts the corresponding smart erasure operations, $\mathcal{E}[p_M]$ and $\mathcal{E}[p_M^\prime]$, wherein the agent parametrizes the erasure protocol for the biased case via an offset $f(\delta)$ that captures the statistical asymmetry. Importantly, this erasure protocol corresponds to a dynamic coupling process to a thermal bath $B$; further details can be found in Refs.~\cite{berut2012experimental, dago2022dynamics}. The final row represents the memory state post-erasure, $\rho_M^\infty$. The minimal entropy cost in each scenario is given by
\begin{equation}
    \mathcal{C}=H ( p_M )-H (\rho_M^\infty) \;\;\;\;\text{and}\;\;\;\;\mathcal{C}^\prime=H ( p_M^\prime )-H (\rho_M^\infty),
\end{equation}
with  $\mathcal{C}^\prime < \mathcal{C}$ reflecting the statistical advantage afforded by the biased distribution.

This example demonstrates that the agent’s knowledge of the memory distribution allows the erasure operation to be optimized, reducing the entropic cost compared with a protocol that ignores the stored information. By tailoring $\mathcal{E}$ to the prior statistics, the agent can attain the minimal entropy cost attainable given the actual uncertainty in the memory, motivating the term \emph{smart} erasure.\\

\section{Diagrammatic View of Entropy and Erasure Costs}\label{supp_mat:venn}

The purpose of this section is to provide a diagrammatic representation of the entropy and information flows underlying the difference between sequential and parallel erasure strategies. We focus on the simplest nontrivial case of repeated measurements, $n=2$, and use entropy Venn diagrams to visualize how correlations between the system and memory registers are structured and processed. Throughout this section, areas represent entropies and overlaps represent mutual informations. For simplicity, we set $H(\rho_M^\infty)=0$.\\

\begin{figure*}[h]
    \centering
    \includegraphics[width=0.5\linewidth]{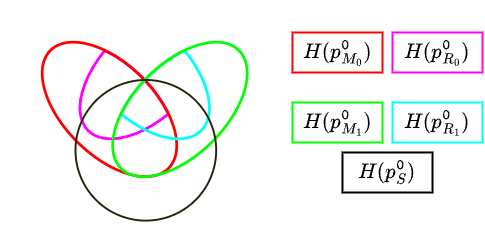}
    \caption{Venn-diagram representation of the joint entropy of a system $S$ and two memories $M_0$ and $M_1$, together with their respective registers $R_0 \subset M_0$ and $R_1 \subset M_1$ used for inference. The area of each region represents the entropy of the corresponding subsystem, while the overlap between any two regions $A$ and $B$ represents their mutual information $I(A : B)$.}
    \label{fig:joint_ent}
\end{figure*}

We consider the joint entropy of the system $S$ and two memories $M_0$ and $M_1$, prior to any erasure, which may be represented diagrammatically as in Fig.~(\ref{fig:joint_ent}). Importantly, Fig.~\ref{fig:joint_ent} represents only the entropy of the joint probability distribution over the states of $(S,M_0,M_1)$. As such, it captures the entropic and informational relationships among these variables, but does not by itself specify how the correlations are implemented physically, nor the temporal order in which they are established.\\

This observation underlies a key point: the same joint-entropy diagram is consistent with both parallel and sequential implementations of the measure–infer–erase cycle, provided one carefully interprets the meaning of the memory labels. In a \emph{parallel} implementation, the subscripts label distinct physical memories. That is, $M_0$ and $M_1$, correspond to two separate memories that are simultaneously correlated with the system $S$. In this case, $R_0$ and $R_1$ reside in different physical devices, and the diagram should be read as describing correlations distributed across multiple subsystems at a single time. In contrast, in a \emph{sequential} implementation, $M_0$ and $M_1$ correspond to the same physical memory at different stages of the protocol. Here, the subscripts do not denote distinct devices, but rather index successive time steps: $M_0$ is the memory state after the first measurement, while $M_1$ is the memory state after the second. The registers $R_0$ and $R_1$ should be understood analogously as the degrees of freedom used for inference at the corresponding iterations.\\

\begin{figure*}[h]
    \centering
    \includegraphics[width=0.8\linewidth]{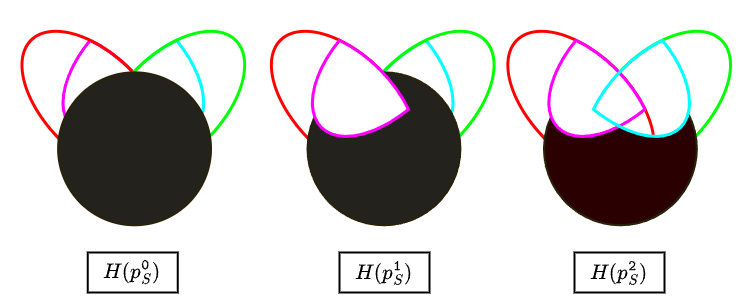}
    \caption{Expected entropy reduction on $S$ through correlations with $R_0$ and $R_1$.}
    \label{fig:venn_inference}
\end{figure*}

Despite these distinct physical interpretations, the joint distribution over $(S,M_0,M_1)$ is identical in the parallel and sequential cases. This equivalence implies that the inference step depends only on the joint statistics of the system and memories, and not on whether information is processed iteratively or jointly. This is illustrated schematically in Fig.~\ref{fig:venn_inference}. In particular, the expected entropy of the system evolves as
\begin{equation}
    H(p_{S}^\mathtt{0}) \;\to\; H(p_{S}^\mathtt{0})-I(p_{S}^\mathtt{0}:p_{R_0}^\mathtt{0}) 
    \;\to\; H(p_{S}^\mathtt{0})-I(p_{S}^\mathtt{0}:p_{R_0}^\mathtt{0})-I(p_{S|R_{0}}^\mathtt{0}:p_{R_1|R_0}^\mathtt{0})
    \;=\; H(p_{S}^\mathtt{0})-I(p_{S}^\mathtt{0}:p_{R_0,R_1}^\mathtt{0}),
\end{equation}
where $I(p_{S|R_{0}}^\mathtt{0}:p_{R_1|R_0}^\mathtt{0})$ should be understood as the mutual information between the distributions of $S$ and $R_1$ conditioned on the state of $R_0$. Notice that the same final reduction in entropy is obtained whether one conditions first on the outcome stored in $R_1$ and then on $R_0$, or incorporates both outcomes simultaneously. This order-independence follows directly from the chain rule
\begin{equation}
    I(p_A:p_B)+I(p_{A|B}:p_{C|B})=I(p_{A}:p_{A,B}),
\end{equation}
and reflects the fact that inference is determined entirely by the joint distribution, independent of the temporal or physical interpretation of the memory registers.\\

\begin{figure*}[h]
    \centering
    \includegraphics[width=0.75\linewidth]{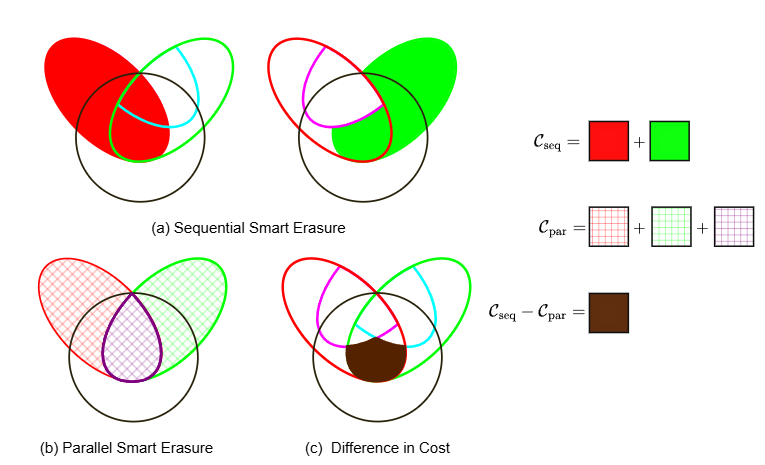}
    \caption{Diagrammatic representation of erasure costs for sequential and parallel protocols. (a) \emph{Sequential erasure}: the shaded regions indicate the minimal cost of erasing each memory in turn, wherein correlations accessible through previously stored registers can be exploited. (b) \emph{Parallel erasure}: when all memories are erased jointly, the full set of correlations across memories can be leveraged; the hatched region denotes the minimal cost in a single smart-erasure operation, leveraging all correlations between the two memories. (c) \emph{Cost difference}: The highlighted region identifies the portion of entropy responsible for the excess cost of the sequential protocol, corresponding to correlations that are present across memories but not accessible through inference registers alone.}
    \label{fig:venn_erasure}
\end{figure*}

Unlike the inference step, the minimal cost of erasure depends explicitly on how correlations are accessed and exploited during the reset operation. This dependence is illustrated by the entropy diagrams in Fig.~\ref{fig:venn_erasure}, which visually accounts for the erasure costs associated with the sequential and parallel protocols.\\

In the sequential protocol, erasure is performed iteratively. The cost of erasing the first memory is simply $H(p_{M_0}^\mathtt{0})$. When erasing the second memory, however, correlations with the previously stored register $R_0$ can be exploited, reducing the cost to the conditional entropy $H(p_{M_1}^\mathtt{1})=H(p_{M_1|R_0}^\mathtt{0})$. The total erasure cost in the sequential protocol is therefore
\begin{equation}
    \mathcal{C}_\text{seq}=H(p_{M_0}^\mathtt{0})+H(p_{M_1|R_0}^\mathtt{0})=H(p_{M_0}^\mathtt{0})+H(p_{M_1}^\mathtt{0})-I(p_{M_1}^\mathtt{0}:p_{R_0}^\mathtt{0}).
\end{equation}
By contrast, in the parallel protocol both memories coexist and are available simultaneously at the time of erasure. In this case, all correlations between $M_0$ and $M_1$ can be leveraged when implementing an optimal erasure operation. The corresponding cost is
\begin{equation}
    \mathcal{C}_\text{par}=H(p_{M_0,M_1}^\mathtt{0})=H(p_{M_0}^\mathtt{0})+H(p_{M_1}^\mathtt{0})-I(p_{M_1}^\mathtt{0}:p_{M_0}^\mathtt{0}).
\end{equation}
The difference between the two protocols is therefore
\begin{equation}
    \mathcal{C}_\text{seq}-\mathcal{C}_\text{par}=I(p_{M_1}^\mathtt{0}:p_{M_0}^\mathtt{0})-I(p_{M_1}^\mathtt{0}:p_{R_0}^\mathtt{0})=I(p_{M_1|R_0}^\mathtt{0}:p_{Q_0|R_0}^\mathtt{0}),
\end{equation}
This quantity measures the correlations between the states $M_1$ and $M_0$ that are not accessible through the register $R_0$ alone. The excess cost of the sequential protocol thus arises not from a lack of correlations, but from constraints on when and how those correlations can be exploited during erasure.\\

\section{Bounds and Formulae for the Example}
\label{supp_mat:example}

In the example discussed in the main text, all entropic quantities are expressed in bits. The memory $M$ is initialized in the state
\begin{equation}
\rho_M^\infty =
\underbrace{ \big( (1-\varepsilon)\dyad{g_c}
+\varepsilon \dyad{g_i} \big)^{\otimes 3} }_{\rho_Q^\infty}
\;\otimes\;
\underbrace{\dyad{\boldsymbol{0}}}_{\rho_R^\infty}.
\end{equation}

\begin{figure*}[h]
    \centering
    \includegraphics[width=0.45\linewidth]{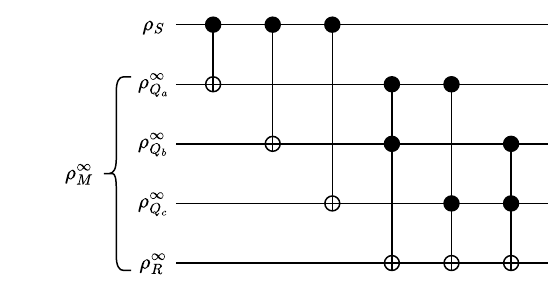}
    \caption{Circuit representation of the measurement operation for the classical bit example. Information is transmitted from the system $S$ to the inaccessible memory $Q$ via CNOT gates, and from $Q$ to the accessible majority-vote register $R$ via CCNOT gates.}
    \label{fig:ExampleMeas}
\end{figure*}

During measurement, $\ket{g_c}$ and $\ket{g_i}$ encode the system state correctly or incorrectly, respectively. This process can be represented in circuit notation, Fig.~(\ref{fig:ExampleMeas}), by mapping $\ket{g_c} \to \ket{0}$ and $\ket{g_i} \to \ket{1}$, using CNOT gates for transmission from $S$ to each bit in $Q$, and CCNOT gates for propagating the majority outcome to $R$. The resulting joint distribution is specified by the likelihoods
\begin{align}
    p(q_\alpha|x) &= 
    \begin{cases}
        1-\varepsilon, & q_\alpha = x,\\
        \varepsilon, & q_\alpha \neq x,
    \end{cases} \\
    p(r|x) &=
    \begin{cases}
        1-\delta, & r = x,\\
        \delta, & r \neq x,
    \end{cases}
\end{align}
where $\varepsilon$ and $\delta$ quantify the noise in the inaccessible memory $Q$ and the accessible register $R$, respectively, with $\delta = \varepsilon^3 + 3\varepsilon^2(1-\varepsilon)$. 

In the absence of an environment $\rho_E$, the relevant single-cycle and cumulative costs reduce to
\begin{align}
    \mathcal{C}_0 &= I(p_{S}^\mathtt{0}:p_{M_0}^\mathtt{0}), \\
    \mathcal{C}_\text{par}(n) &= I(p_{S}^\mathtt{0}:p_{M_{0::n-1}}^\mathtt{0}), \\
    \mathcal{C}_\text{seq}(n) &= \mathcal{C}_0+\sum_{k=1}^{n-1} I(p_{S|R_{0::k-1}}^\mathtt{0}:p_{M_{k}|R_{0::k-1}}^\mathtt{0}).
\end{align}
Although Fig.~(\ref{fig:ex}) is based on numerical evaluation of the above equations, we derive analytical bounds and behaviors of the information gain and erasure costs can be derived explicitly.

\subsubsection{Information Gain}
To characterize the performance for $n$ repeated measurements, we require bounds on the total information gained, $\mathcal{I}(n)=I(p_{S}^\mathtt{0}:p_{R_{0::n-1}}^\mathtt{0})$, where the states of each register $R_k$ can be mapped to an independent binary random variable with error probability $\delta$. To compute a lower bound, we make use of the following inequality for the binary entropy function
\begin{equation}
    h(q) \leq 2 \sqrt{q(1-q)},
\end{equation}
from which we obtain
\begin{equation}
\begin{split}
    \mathcal{I}(n) &=I(p_{S}^\mathtt{0}:p_{R_{0::n-1}}^\mathtt{0}) \\
     &=H(p_S^\mathtt{0})-H(p_{S|R_{0::n-1}}^\mathtt{0}) \\
    &=1-\sum_{r_{0::n-1}}p(r_{0::n-1})h(p(x=0|r_{0::n-1})) \\
    &\geq 1-2\sum_{r_{0::n-1}}p(r_{0::n-1})\sqrt{p(x=0|r_{0::n-1})p(x=1|r_{0::n-1})} \\
    &= 1-\sum_{k=0}^n \binom{n}{k} \big( \delta (1-\delta) \big)^{n/2} \\
    &= 1- \big[ 4 \delta (1-\delta) \big]^{n/2}.
\end{split}
\end{equation}
For an upper bound, we make use of the concavity of the logarithm
\begin{equation}
\begin{split}
    \mathcal{I}(n) &= \sum_{x,r_{0::n-1}} p(r_{0::n-1}|x) p_S^\mathtt{0}(x) \log_2 \frac{p(r_{0::n-1}|x)  }{p_{R_{0::n-1}}^\mathtt{0}(r_{0::n-1}) }\\
    &\leq \log_2 \Big( \sum_{x,r_{0::n-1}} \frac{p(r_{0::n-1}|x)^2 p_S^\mathtt{0}(x)  }{p_{R_{0::n-1}}^\mathtt{0}(r_{0::n-1}) }  \Big)  \\
    &= \log_2 \Big( \sum_{r_{0::n-1}} \frac{p(r_{0::n-1}|x=0)^2+p(r_{0::n-1}|x=1)^2  }{p(r_{0::n-1}|x=0)+p(r_{0::n-1}|x=1) }  \Big) \\
    &= \log_2 \Big( 2-2\sum_{r_{0::n-1}} \frac{p(r_{0::n-1}|x=0)p(r_{0::n-1}|x=1)  }{p(r_{0::n-1}|x=0)+p(r_{0::n-1}|x=1) }  \Big) \\
    &= \log_2 \Big( 2-2\sum_{k=0}^{n}  \binom{n}{k}\frac{\big( \delta (1-\delta) \big)^n }{\delta^k (1-\delta)^{n-k}+\delta^{n-k}(1-\delta)^{k}}  \Big) \\
    &\leq \log_2 \big[ 2- \big(4\delta(1-\delta)  \big)^n \big].
\end{split}
\end{equation}
Thus the information gain satisfies
\begin{equation}
\label{eq:info_gain_ex_bound}
    \log_2 \big(2-\big[4 \delta(1-\delta) \big]^n \big) \geq \mathcal{I}(n) \geq 1- \big[4 \delta(1-\delta) \big]^{n/2}.
\end{equation}

\subsubsection{Parallel Cost}
Because the parallel cost, $\mathcal{C}_\text{par}(n)=I(p_{S}^\mathtt{0}:p_{M_{0::n-1}}^\mathtt{0})$, has a very similar structure to the information gain in this example, the same bounding techniques can be applied. An analogous expression to Eq.~\eqref{eq:info_gain_ex_bound} is obtained with $\delta \to \varepsilon$ and $n \to 3n$.

\subsubsection{Sequential Cost}
For the specific memory structure considered in this example, we have $H(p^0_{M_k})=H(p^0_{Q_k})$. As a result, the difference
between the parallel and sequential costs can be written as
\begin{equation}
\begin{split}
    \mathcal{C}_\text{seq}(n)-\mathcal{C}_\text{par}(n) &=\sum_{k=1}^{n-1} \big( H(p^\mathtt{0}_{M_k|R_{0::k-1}})-H(p^\mathtt{0}_{M_k|M_{0::k-1}}) \big) \\
    &= \sum_{k=1}^{n-1} \big( H(p^\mathtt{0}_{Q_k|R_{0::k-1}})-H(p^\mathtt{0}_{Q_k|Q_{0::k-1}}) \big)\\
    &=\sum_{k=1}^{n-1} \sum_{q_{0::k-1}}p^\mathtt{0}_{Q_{0::k-1}}(q_{0::k-1}) \sum_{q_k} p(q_k|q_{0::k-1}) \log_2 \frac{p(q_k|q_{0::k-1})}{p(q_k|r_{0::k-1})} \\
    &=\sum_{k=1}^{n-1} \langle D_{KL} \rangle^{(k)}.
\end{split}
\end{equation}
For each $k$, the inner expression
\begin{equation}
    \langle D_{KL} \rangle^{(k)}=\sum_{q_{0::k-1}}p^\mathtt{0}_{Q_{0::k-1}}(q_{0::k-1}) \sum_{q_k} p(q_k|q_{0::k-1}) \log_2 \frac{p(q_k|q_{0::k-1})}{p(q_k|r_{0::k-1})}
\end{equation}
is precisely the average Kullback-Leibler divergence between the conditional distribution of $Q_k$ given $q_{0::k-1}$ and the conditional distribution of $Q_k$ given $r_{0::k-1}$. Summing over all $2^{3k}$ possible states of $Q_{0::k-1}$ can be done systematically: denote the number of $R$ registers with state 0 by $j$, and count error terms in $Q$ with $a$ for type 001/010/100 and $b$ for type 110/101/011, then one can map
\begin{equation}
    p_{Q_{0::k-1}}^\mathtt{0}(q_{0::k-1}) \to \frac{1}{2} \big( (1-\varepsilon)^{3j-a+b}\varepsilon^{a-b+3(k-j)}+\varepsilon^{3j-a+b}(1-\varepsilon)^{a-b+3(k-j)} \big),
\end{equation}
and similarly
\begin{equation}
    p_{R_{0::k-1}}^\mathtt{0}(r_{0::k-1}) \to \frac{1}{2} \big( (1-\delta)^{j}\delta^{(k-j)}+\delta^{j}(1-\delta)^{k-j} \big).
\end{equation}

For generic values of $\varepsilon$, the quantity $\langle D_{KL} \rangle^{(k)}$ does not admit a closed-form expression. Nevertheless, its behaviour can be characterized accurately in asymptotic regimes by means of series expansions. Owing to the symmetry of the underlying distributions, it is convenient to parameterize these expansions in terms of $\alpha = 4\varepsilon(1-\varepsilon)$, which interpolates between the deterministic ($\alpha \to 0$) and maximally noisy ($\alpha \to 1$) regimes. In the deterministic regime, only the first two terms contribute up to third order in $\alpha$, yielding
\begin{equation}
\begin{split}
    \mathcal{C}_\text{seq}(n)-\mathcal{C}_\text{par}(n) & =  \langle D_{KL} \rangle^{(1)}+\langle D_{KL} \rangle^{(2)} + \mathcal{O} \big( \alpha^4 \big) \\
    &= \frac{3\alpha^2}{8} \log_2 \frac{16}{3\alpha}-\frac{\alpha^3}{8} \big( \frac{19}{8\ln2}+\log_2\frac{4}{3\alpha^2} \big) + \mathcal{O} \big( \alpha^4 \big).
\end{split}
\end{equation}
By contrast, in the maximally noisy regime, contributions from all $k$ become relevant, yielding
\begin{equation}
\begin{split}
    \mathcal{C}_\text{seq}(n)-\mathcal{C}_\text{par}(n) & =  \sum_{k=1}^{n-1}\langle D_{KL} \rangle^{(k)} \\
    & = \frac{9}{4\ln2} \sum_{k=1}^{n-1} \Big(k(1-\alpha)^2+\frac{k(21k+11)}{4}(1-\alpha)^3 +\mathcal{O} \big( k^3 (1-\alpha)^4 \big) \Big) \\
    &=  \frac{9n(n-1)}{16\ln2}(1-\alpha)^2 \big(2+(7n+2)(1-\alpha) \big) +\mathcal{O} \big( n^4 (1-\alpha)^4 \big) \Big).
\end{split}
\end{equation}

\end{document}